\documentclass[nofootinbib,aps]{revtex4}

\usepackage{amstext,amsmath,amssymb}
\usepackage[dvips]{graphicx}
\usepackage{latexsym}

\newcommand{\Ref}[1]{(\ref{#1})}

\newcommand{\R}{\mathbb{R}}
\newcommand{\C}{\mathbb{C}}

\def\be{\begin{equation}}
\def\ee{\end{equation}}
\def\bes{\begin{eqnarray}}
\def\ees{\end{eqnarray}}
\def\nn{\nonumber}
\def\arr{\rightarrow}

\def\la{\langle}
\def\ra{\rangle}
\def\f{\frac}

\def \vphi{\varphi}
\newcommand{\SU}{\mathrm{SU}}
\newcommand{\SL}{\mathrm{SL}}
\newcommand{\SO}{\mathrm{SO}}

\def\pp{\partial}
\def\hh{{\cal H}}
\def\kk{{\cal K}}
\def\pp{{\cal P}}
\def\cc{{\cal C}}
\def\ii{{\cal I}}
\def\vv{{\cal V}}
\def\aa{{\cal A}}

\begin{document}

\title{Coupling of spacetime atoms and spin foam renormalisation from group field theory}

\author{{\bf Etera R. Livine}\footnote{elivine@perimeterinstitute.ca}}
\affiliation{Perimeter Institute, 31 Caroline Street North
Waterloo, Ontario, Canada N2L 2Y5}
\affiliation{Laboratoire de Physique, ENS Lyon, CNRS UMR 5672, 46 All\'ee d'Italie, 69364 Lyon Cedex 07}
\author{{\bf Daniele Oriti}\footnote{d.oriti@damtp.cam.ac.uk}}
\affiliation{Department of Applied Mathematics and Theoretical Physics,
Centre for Mathematical Sciences, University of Cambridge,
Wilberforce Road, Cambridge CB3 0WA, UK,EU,}
\affiliation{Girton College, University of Cambridge, Cambridge CB3 0JG, UK, EU}

\begin{abstract}

\begin{center}
{\small ABSTRACT}
\end{center}
We study the issue of coupling among 4-simplices in the context of spin foam models obtained from a group field theory formalism. We construct a generalisation of the Barrett-Crane model in which an additional coupling between the normals to tetrahedra, as defined in different 4-simplices that share them, is present. This is realised through an extension of the usual field over the group manifold to a five argument one. We define a specific model in which this coupling is parametrised by an additional real parameter that allows to tune the degree of locality of the resulting model, interpolating between the usual Barrett-Crane model and a flat BF-type one. Moreover, we define a further extension of the group field theory formalism in which the coupling parameter enters as a new variable of the field, and the action presents derivative terms that lead to modified classical equations of motion. Finally, we discuss the issue of renormalisation of spin foam models, and how the new coupled model can be of help regarding this. 

\end{abstract}

\maketitle

\tableofcontents

\section{Introduction}

There is no established complete theory of 4-dimensional quantum gravity as yet, and several approaches are being pursued to achieve this goal; among these, spin foam models \cite{review, alexreview} represent a new promising one in that they seem to be a point of convergence of quite a few lines of research, including loop quantum gravity and simplicial approaches. Among the various models proposed, the Barrett-Crane model \cite{bc} is the one on which most of the work has focused so far; the basic idea of this model is to describe and quantize gravity, in a simplicial setting, as a constrained topological field theory, where the topological field theory one starts with is the so-called BF theory. The result is a purely combinatorial and algebraic version of a sum-over-histories formulation of the theory in which spacetime is replaced by a combinatorial 2-complex and spacetime geometry is encoded in algebraic data from the representation theory of the Lorentz group. These data are of two types: representation parameters labelling (unitary irreducible) simple, i.e. class I, representations of the group (in the principal series), and vectors in the homogeneous space ${\cal H}^3 = \SO(3,1)/\SO(3)$; both have a geometric interpretation within the simplicial spacetime one reconstructs from the spin foam (see \cite{us} and references therein): the first are interpreted as areas of the triangles they label, while the second are thought of as normals to the tetrahedra of the simplicial complex. In this simplicial context, the spacetime behind spin foam models is made out of fundamental building blocks, \lq\lq atoms of a quantum spacetime\rq\rq, being the 4-simplices of the triangulation, or equivalently, the vertices of the spin foam, and the spin foam model describe how these building blocks interact, how they are coupled, with this coupling determining the details of the overall spacetime geometry. In the specific case of the Barrett-Crane model the coupling between 4-simplices is limited to the first set of algebraic data, i.e. to the representation labels only, in that triangles have the same area in all the 4-simplices they are shared by; on the other hand, there two normal variables for each tetrahedron in the manifold, one for each 4-simplex sharing it, and these two variables are completely un-correlated, i.e. un-coupled. This is not necessarily a problem and in fact it has a nice geometric interpretation (see \cite{us} and references therein), but it represents a sort of \lq\lq ultra-locality\rq\rq of the model that one may want to relax (on the ultra-locality of the Barrett-Crane model seen from the point of view of group field theories, see \cite{generalised}). One reason for this is that one expects this coupling to be affected by and somehow flow under renormalisation of the quantum amplitudes, and one would like to be able to tune the model to different degrees of locality, so to allow for propagating degrees of freedom; in fact it is not clear whether or not these are to be expected in a ultra-local model.
Therefore, one would like to introduce an additional coupling between 4-simplices with respect to the Barrett-Crane model, and possibly in a tunable way, e.g. introducing an extra parameter in the theory that can be set, when wanted, to a value that gives back the un-coupled Barrett-Crane model, and on whose value one can play to obtain the various levels of locality. This is exactly what we achieve in this paper. The goal goes however beyond the construction of a coupled model: one would like to study the renormalisation of the coupling, as a step towards the difficult task of understanding the renormalisation of spin foam models in full generality, an issue made tricky by the absence of any background structure, as it should be the case for any sensible candidate to a theory of quantum gravity. We believe the model presented here may be of help regarding this more ambitious goal.
In this article we work in the framework and language of group field theory \cite{review, alexreview,laurentgft}. Group field theories are known to be behind any spin foam model we know so far \cite{carlomike} and to give the most complete formulation of them, in that they allow to remove the dependence on any specific discretization of spacetime obtaining thus a fully background independent model. Moreover, group field theories allow for an easy control over the degrees of freedom and the variables being dealt with in that it provides the basic ingredients of the spin foam model without being tied to any choice of underlying 2-complex, because the provide directly the basic building blocks of it, in the form of propagators and vertex amplitudes forming it as a Feynman diagram of the theory; it is then easier to work in the group field theory context when looking for generalisations of the current models because one sees clearly the structures being generalised even before one sees how the resulting spin foam amplitudes end up being modified. The use of the group field theory formalism descends however also, independently on the specific result one wants to achieve, from the general point of view that group field theories represent not only the most complete formulation of spin foam models, as said, but the most {\it fundamental} definition of them, i.e. they are {\it the} theory we are talking about when discussing spin foam models, whose content and significance goes far beyond that of spin foam models alone. This point of view is not that outrageous given that indeed spin foam models arise as Feynman diagrams amplitudes for the corresponding group field theory, and we know that there is much  more in a quantum field theory than its perturbative expansion in Feynman diagrams. However, at present there are too many aspects of group field theories that are not been studied and too many properties of spin foam models whose group field theory origin is not understood for being able to consider this point of view as {\it the correct one} and not just a possible one. We content ourselves in this paper to show one instance in which working directly at the group field theory level proves advantageous over working at the level of its Feynman amplitudes. Moreover, the results we will present in the end of this work, concerning a new type of group field theory suggest that the group field theory formalism may contain more than usually suspected also with respect to the issue of renormalisation of spin foam models. 

The paper is organised as follows: we start by showing how the {\it usual} Barrett-Crane model can be obtained by a group field theory in for a field with {\it five} arguments living in the group manifold, with the extra (with respect to the usual formulations) fifth variable playing the role, in the resulting amplitudes, of the normal to the tetrahedra of the triangulation dual to the Feynman 2-complex, but with the corresponding degrees of freedom not being propagated from vertex to vertex (from 4-simplex to 4-simplex); then in section \ref{sec:genmodel} we construct a generalisation of this model that provides a non-trivial coupling of these variables across different 4-simplices; we write down the model in full generality, i.e. regardless of any specific choice of coupling function, but we then exhibit a particularly interesting example, with the coupling function being the heat kernel on the group, and thus carrying a tunable coupling parameter, and leading to a model interpolating between the Barrett-Crane model and a strongly coupled flat model, as we are going to discuss; in section \ref{sec:partBC} we discuss how the mathematical structures involved in this generalised model are relevant also for the insertion of particles in the Barrett-Crane model; finally, in section \ref{sec:gengft} we propose a further generalisation of the group field theory formalism in which the coupling parameter of the coupled model is promoted to a variable on the same footing as group variables, and in section \ref{sec: renorm} we discuss at length the issue of renormalisation of spin foam model from the novel perspective provided by the results of this work.             

\section{Generalized Barrett-Crane model and the coupling of spacetime
atoms} \label{sec:genBC}
\subsection{Cosmetics on the Barrett-Crane model} \label{sec:cosm}
Before we show how it is possible to generalise the Barrett-Crane model to a new model that presents a non-trivial coupling between 4-simplices, we want to first show how the {\it usual} Barrett-Crane model can be derived from a group field theory action defined for a field living on {\it five} copies of the Lorentz group, instead of four, with the fifth group element having the interpretation of the normal to the tetrahedron of which the field represents the (second) quantization.

We start with a complex field $\vphi(g_i,G)$\footnote{Here and in the following we use a simplified notation labelling by a single group element $g_i$, indexed by $i$, the four group elements $g_1,g_2,g_3,g_4$ variables of the field} on $\SO(3,1)^{\times 5}$; this is our dynamical object\footnote{The calculations and results of this paper are valid for both $\SO(3,1)$ and $\SO(4)$ in the same way, with the non-compactness of the Lorentz group causing only a higher degree of formal complications but no difference in substance; therefore in what follows we always refer explicitly only to $\SO(3,1)$ because it is physically more interesting, but we use the notation that is customary in the $\SO(4)$ case instead because it is easier to follow and less cumbersome.}. We define its projection on $(\SO(3,1)/\SO(3))^{\times 4}\times \SO(3,1)$ and expand it into Fourier modes:
$$
P_h\vphi(g_i,G)=\sum_{J_i A_i}\vphi^{J_i}_{A_i,B_i}(G)D^{J_i}_{A_iB_i}(g_i)W^{J_i}_{B_i},
$$
where $\vphi^{J_i}_{A_i}(G)$ are the Fourier modes of the field and $W^{J_i}_{A_i}$ the $\SO(3)$-invariant vector in the $J_i$ (simple) representation.
The gauge invariant field is defined as:
\bes
\phi(g_i,G)\equiv P_gP_h\vphi(g_i,G)
=\int d\Lambda\, P_h\vphi(g_i\Lambda,G\Lambda) \\ = \int d\Lambda\, \sum_{J_i A_i}\vphi^{J_i}_{A_i}(G\Lambda)D^{J_i}_{A_iB_i}(g_i\Lambda)W^{J_i}_{B_i} = \label{eq:field1} \\
=\int d\Lambda\, \sum_{J_i A_i}\vphi^{J_i}_{A_i}(\Lambda)D^{J_i}_{A_iB_i}(g_iG^{-1}\Lambda)W^{J_i}_{B_i}.
\ees
The group variable $\Lambda$ is then naturally projected down to the hyperboloid $\hh=\SO(3,1)/\SO(3)$.
The field is a function of five variables, either five group elements or five representations of the Lorentz group, in the complete mode expansion, so it represents a generalisation of the field on which the Barrett-Crane model is based; this is also clear if one thinks that it decomposes into combinations of 5-valent intertwiners between simple representations. In a sense it can be said that it \lq\lq unlocks\rq\rq the structure of a quantum tetrahedron, by relaxing the closure constraint and allowing the four simple representations corresponding to the 4 triangles of a tetrahedron to be mapped to a non-trivial one. However, whether and how this is realised depends on what action one defines for this extended field, i.e. on whether or not one chooses to trivialise the dependence on the extra variables. Moreover, we see that for each given value of the $G$ variables, the extra fifth variables of the field are given by the $\Lambda$s entering the amplitudes as normals to the tetrahedra, as anticipated. The presence of these variables, however, allows for $\Lambda$s to be coupled in different 4-simplices, as we will see.  

In fact, using this definition of the field, one can define the action:
\bes 
S[\vphi]&=& \frac{1}{2}\prod_i\int dg_i \int dG\int dG'\phi(g_i,G)\phi(g_i,G') \\
&+&\frac{\lambda}{5!} \prod_{i=1}^{10} \int dg_i\,\int dG_A ..dG_E
\phi(g_1,g_2,g_3,g_4,G_A)\phi(g_4,g_5,g_6,g_7,G_B)\phi(g_7,..,G_C)
\phi(g_9,..,G_D)\phi(g_{10},..,G_E), \nn
\ees
with capital letters labeling the five group elements associated to the five tetrahedra of a 4-simplex.
This is a trivial extension of the group field theory action from which the usual Barrett-Crane model is derived, in the formulation of \cite{DP-F-K-R}, in the sense that the above action reproduces {\it exactly} the same amplitudes of the usual Barrett-Crane model, as it is easy to verify. For example, we can show how this is realised for the propagator of the theory. The kinetic term is decomposed in modes by standard methods, starting from expression \Ref{eq:field1} as:
\be
\prod_i\int dg_i \int dG  dG' \phi(g_i,G)\phi(g_i,G')= \int d\Lambda d\Lambda' \int dG dG' \vphi^{J_i,J}_{A_i,m,n} \vphi^{J_i,J'}_{A_i,k,l} \prod_i D^{J_i}_{00}(\Lambda \Lambda'^{-1})\, D^{J}_{mn}(G\Lambda)\,D^{J'}_{kl}(G'\Lambda'),
\ee
with implicit summation over all indices $J_i, A_i, J,J',m,n,k,l$. It is easy to see that, upon integration of the $G$ variables, the dependence on the variables $\Lambda$ (those imposing the gauge invariance of the fields) is reduced and the fifth representations $J$,$J'$ are forced to be the singlet, and the usual kinetic term \cite{DP-F-K-R} is obtained:
\be
\prod_i\int dg_i \int dG \int dG' \phi(g_i,G)\phi(g_i,G')= \sum_{J_i,A_i} \vphi^{J_i}_{A_i}\vphi^{J_i}_{A_i}\int d\Lambda d\Lambda' \prod_i D^{J_i}_{00}(\Lambda \Lambda'^{-1}). 
\ee
The same can be verified for the interaction term, that gives the usual Barrett-Crane amplitude for a 4-simplex. In the end the Barrett-Crane model is obtained also in this 5-arguments formalism, with the (once) fifth variable entering in the model as the normal to the tetrahedra in each of the 4-simplices to which they belong \cite{us}, but being completely decoupled in the different 4-simplices. Note that, had we imposed the projection $P_h$ only in the interaction term and not in the kinetic one, we would have obtained back the Barrett-Crane model in the version given in \cite{P-R}; so the same relation between the various versions of the BC model is present in this five argument extension as in the usual four argument formulation of the group field theory. In the following we will discuss in parallel what happens when one does or does not impose the simplicity constraints in the kinetic term.
The usefulness of this extension of the model lies in the fact that, the dependence on the normal variables having been made explicit in the field theory, it is possible to modify the field theory itself in a very simple way, and thus to introduce a coupling between these variables. This is what we are now going to show.

\subsection{A generalised model with a non-trivial coupling between 4-simplices} \label{sec:genmodel}
We modify the kinetic term, the one producing the propagator of the theory and thus the edge amplitudes of the spin foam model, in the above extended action by inserting an arbitrary coupling for the new extra variables $G$:
\be
S_{\rm kin}[\vphi]\equiv
\int dg_i \int dG \, dG'\,
f(G,G')\phi(g_i,G)\phi(g_i,G').
\ee
Thus the function $f(G,G')$ encodes in configuration space a non-trivial coupling between the fifth variables of the field, that, as we have seen in the previous section, enter in each 4-simplex (vertex) amplitude as the normals to the tetrahedra in the reference frame associated to each of them. It represents a non-trivial amplitude for the curvature, in the sense that it does not weight equally all the possible values of them, as it does the trivial choice $f(G,G')=1$. The coupling $f$ is naturally of the form $f(G,G')\equiv F(G'G^{-1})$ where $F$ is a function over $\SO(3,1)$.
Moreover, we would like to have a Lorentz invariant coupling such that the correlations be invariant under $\Lambda,\Lambda'\arr g\Lambda,g\Lambda'$ for all $g\in\SO(3,1)$. This amounts to assuming the coupling function $F$ to be invariant under conjugation: $F(G)=F(g^{-1}Gg), \,\forall g$.

We want now to see what expression this coupling takes in momentum (representation) space; this is necessary when the simplicity constraints are imposed, i.e. in trying to construct a \lq\lq coupled extension\rq\rq of the version of the Barrett-Crane model presented in \cite{DP-F-K-R}, because when this is done taking the inverse of the kinetic operator in configuration space is particularly cumbersome, and much simpler in momentum space. Expanding into Fourier modes, we obtain:
\be
S_{\rm kin}[\vphi]=
\int d\Lambda\, d\Lambda'\, \sum_{J_i,A_i}\f{1}{\Delta_{J_i}}
\vphi^{J_i}_{A_i}(\Lambda)\vphi^{J_i}_{A_i}(\Lambda')
\,\left[
\int dG\,dG'\, f(G,G') \prod_{i=1}^4 K^{J_i}(\Lambda'^{-1}G'G^{-1}\Lambda)
\right],
\ee
where $K^{J}(g)=\bar{W}^J_AD^J_{AB}(g)W^J_B=D^{J}_{00}(g)$ is the homogeneous kernel, the Hadamard propagator of 'mass' $J$ (more precisely, $m^2=J(J+1)$) on the hyperboloid $\hh$ \cite{feynman}.
Then the quadratic coupling in the action is given by a kind of Fourier transform of $F$ where the modes are the \lq\lq eye diagram evaluations\rq\rq:
\be
{\cal C}^{J_i}(\Lambda,\Lambda')\equiv
\int dG\, F(G) \prod_{i=1}^4 K^{J_i}(\Lambda'^{-1}G\Lambda).
\label{transform}
\ee
Because $F$ is invariant under conjugation, ${\cal C}^{J_i}$ is simply
a function of $\Lambda'^{-1}\Lambda$, so we will denote it more concisely
${\cal C}^{J_i}(\Lambda,\Lambda')\equiv {\cal
  C}^{J_i}(\Lambda'^{-1}\Lambda)$, with:
$$
{\cal C}^{J_i}(g)=\int dG\, F(G) \prod_{i=1}^4 K^{J_i}(Gg).
$$
${\cal C}^{J_i}$ is thus a zonal function, i.e. invariant under both
the left and right actions of $\SO(3)$. Then, under the assumption
that it is a $L^2$ function over the hyperboloid, it can be decomposed
into Fourier modes:
$$
{\cal C}^{J_i}(g)=\sum_{J}\Delta_J{\cal C}^{J_i}_J K_J(g),
$$
where ${\cal C}^{J_i}_J$ are its Fourier components.

Let's point out that the modes of the transform \Ref{transform} that
we introduced above $\prod_{i=1}^4 K^{J_i}(\Lambda'^{-1}G\Lambda)$
were already discussed in \cite{us}. Mathematically, it is the
evaluation of the eye diagram labeled by the four representations
$J_i$ on the group element $G$. It can be interpreted as the
probability amplitude of the quantum tetrahedron, with triangles
defined by the four representations $J_i$, carrying a curvature
defined by the parallel transport $G$ between the two 4-simplices
sharing that tetrahedron.

The propagator of the theory will be the inverse kernel
$P(\Lambda,\Lambda')$ such that:
$$
\int_{{\cal H}} d\Lambda'\,
\pp^{J_i}(\Lambda,\Lambda'){\cal
  C}^{J_i}(\Lambda',\Lambda'')=\delta_{{\cal
    H}}(\Lambda\Lambda''^{-1}),
$$
or equivalently
\be
\int_{\SO(3,1)} dg\,
\pp^{J_i}(Gg^{-1}){\cal C}^{J_i}(g)=\delta(G).
\ee
Then considering that $\delta(G)=\sum_J\Delta_JK^J(G)$, it is straightforward to check that:
$$
\pp^{J_i}(g)=\sum_{J}\Delta_J\f{1}{{\cal C}^{J_i}_J} K_J(g).
$$

It is easy to see what changes in the above results if we do not impose the simplicity constraints in the kinetic term of our generalised group field theory action; of course there will still be a non-trivial coupling for the fifth variables of the field, resulting from a non-trivial coupling insertion $f(G,G')$ in configuration space, and the corresponding propagator in configuration space will be given by a product of delta functions for the first four arguments of the field, just as in \cite{P-R} times a propagator for the fifth pair of variable given simply by $f^{-1}(G,G')$. The four deltas and the extra propagator will be intertwined by a global action of $\SO(3,1)$; in momentum space this will lead for the kinetic term to an analogue of the operator $\mathcal{C}^{J_{i}}$, given by:
\be
{\cal C}_{B_i C_i}^{J_i}(g)=\int dG\, F(G) \prod_{i=1}^4 D_{B_i C_i}^{J_i}(Gg).
\ee
with extra indices $B_i,C_i$ (appearing also in the field modes, resulting from the fact that we do not project anymore on the invariant vectors in the relevant representation spaces. The rest of the construction proceeds analogously, leading to a propagator that carries also these extra indices (of course all the variables are in this case in the group and not in the homogeneous space).   

What we have done up to now is completely general, in that it holds for any choice of coupling function $f$, and it gives a generalisation of the Barrett-Crane model with a non-trivial coupling between 4-simplices with respect to the normal vectors variables. 
However, we can do more than this: we can exhibit a specific choice of the coupling function satisfying all the properties required, thus providing us with a specific model with many interesting properties, that we are going to highlight. Moreover, this model leads naturally to a further generalisation of the Barrett-Crane model, to be discussed below, that can be the starting point for a new approach to the issue of renormalisation of spin foam models.
  
Consider the case in which the simplicity projections $P_h$ are not imposed in the kinetic term, as we assume from now on. The specific model we mentioned arise from choosing the (truncated) heat kernel function as the coupling function $f(G,G')$ in configuration space:
\be
f(G,G')=F(g)=\kk^{-\beta,L}(g)=\sum_{J\le L} \Delta_J e^{\beta C_J} K^J(g),
\ee
where $C_J$ is the quadratic Casimir of the Lorentz group; then the propagator is the inverse heat kernel $\kk^{\beta,L}$, i.e. the (truncated) heat kernel function for the opposite value of the coupling $\beta$, as it is easy to verify.
Considering (formally) the heat kernel $\kk^{-\beta}(g)$ (untruncated) as coupling
function $f(G,G')$, two limits are particularly interesting for the corresponding propagator $\kk^{\beta}(g)$:
\begin{itemize}
\item $\beta\arr \infty$: the coupling becomes trivial, i.e.
\be
\kk^{\beta}(g)\rightarrow Id
\qquad \Rightarrow \qquad
C^{J_i}(\Lambda, \Lambda')=1,
\ee
and we lose the
  dependence on $\Lambda$ and $\Lambda'$. We recover the usual expression for the Barrett-Crane model in the expansion in Feynman graphs; in this limit the 4-simplices are decoupled as far as the $\Lambda$ variables are concerned, meaning that the connection, that maps the normal vector to the tetrahedron as seen from the reference frame associated to one 4-simplex to that corresponding to the reference frame associated to the other 4-simplex, is completely arbitrary, and the only coupling is given by the representations assigned to the triangles of the tetrahedron, which are the same in both 4-simplices that share it.
\item $\beta\arr 0$: the coupling becomes rigid, i.e. $\kk^{\beta}(g)=\delta(g)$ so we are in
  a strong coupling limit. The propagator constraints
  $G=G'$, where $G$ and $G'$ are the fifth variables in the two fields corresponding to the same tetrahedron and interacting in the two vertices, being propagated by $\kk^\beta(g)$. We call this model the flat model since it
  amounts to imposing a trivial parallel transport between
  4-simplices, i.e. constraining the (boost part of the) connection to be flat. 
  Here we stress that we constrain the fifth group variables $G, G'$ to be equal and not the normals $\Lambda, \Lambda'$, so the coupling is simply the eye diagram evaluation:
$$
C^{J_i}(\Lambda, \Lambda')=\prod_{i=1}^4K^{J_i}(\Lambda'^{-1}\Lambda).
$$
it would be interesting to understand the explicit relation between this limit of the model and BF theories. On the one hand it would seem that this limit indeed corresponds to a BF theory, since we constrain the curvature to be flat. On the other hand, it is not the BF theory based on the Lorentz group (the Crane-Yetter model), however it could well be a BF theory on the homogeneous space $\SL(2,\C)/SU(2)$.
\end{itemize}
Both these limits can be easily verified both in configuration space and in momentum space, by a straightforward calculation (in this context without simplicity projectors).

Now let us have a look at the interaction term of the group field
theory, which defines the vertex amplitude of a spin foam or
equivalently the 4-simplex amplitude. Let us consider the following
integral:
\be
I(G_A,..,G_E)\equiv
\int \prod_{i=1}^{10} dg_i\,
\phi(g_1,g_2,g_3,g_4,G_A)\phi(g_4,g_5,g_6,g_7,G_B)\phi(g_7,g_3,g_8,g_9,G_C)
\phi(g_9,..,G_D)\phi(g_{10},..,G_E),
\ee
where $i=1,..10$ refers to the triangles of the 4-simplex and  the letters
$A,..,E$ to its tetrahedra.
Then it is straightforward to check that:
\be
I(G_A,..,G_E)=
\int d\Lambda_A\, ..\, d\Lambda_E\,
\sum_{J_i, A_i} \f{1}{\Delta_{J_i}}
\vphi^{J_1..J_4}_{A_1..A_4}(\Lambda_A)..\vphi^{J_{10}..J_1}_{A_{10}..A_1}(\Lambda_E)
\,\prod_{i=1}^{10} K^{J_i}(\Lambda_{t(i)}^{-1}G_{t(i)}G_{s(i)}^{-1}\Lambda_{s(i)}),
\ee
where we label $s(i)$ and $t(i)$ the source and target vertices of the edge $i$ of the 4-simplex spin network \cite{F-K}, or equivalently, the two tetrahedra sharing the triangle $i$ of the 4-simplex.

If we take as interaction term $\int dG_A\, ..\,dG_E I(G_A,..,G_E)$,
then we erase all information on the normals $\Lambda_A,..\Lambda_E$
and we recover the usual Barrett-Crane evaluation of a 4-simplex, as we have discussed in the previous section. As
we would like to keep a non-trivial dependence  of the $\Lambda$'s, we
simply choose:
\be
S_{\rm int}[\vphi]=I(G_A={\rm Id},..,G_E={\rm Id}).
\ee

This way, we keep an extended dependence on the variables $\Lambda$, that can be coupled by the propagator in each Feynman graph, leading to a spin foam model which differs from the Barrett-Crane model. Note that choosing ${\rm Id}$ is not essential. Fixing the $G_v$ variables to another value would amount to changing the origin -the reference point- of the hyperboloid for each variable $\Lambda_v$.

Finally, our group field theory for the field $\vphi$ is defined by the action
$S_{\rm GFT}[\vphi]\equiv S_{\rm kin}[\vphi]+S_{\rm int}[\vphi]$, i.e. by:
\bes
S_{\rm GFT}[\vphi]&\equiv& \int dg_i \int dG \, dG'\,
f(G,G')\phi(g_i,G)\phi(g_i,G')  \nn\\
&&+\frac{\lambda}{5!}  \int \prod_{i=1}^{10} dg_i\,
\phi(g_1,g_2,g_3,g_4,Id)\phi(g_4,g_5,g_6,g_7,Id)\phi(g_7,g_3,g_8,g_9,Id)
\phi(g_9,..,Id)\phi(g_{10},..,Id),
\ees
in configuration space, or
\bes
S_{\rm GFT}[\vphi]&\equiv& \int d\Lambda d\Lambda' \sum_{J_i,A_i}\f{1}{\Delta_{J_i}}
\vphi^{J_i}_{A_i B_i}(\Lambda)\vphi^{J_i}_{A_i C_i}(\Lambda'){\cal C}_{B_i C_i}^{J_i}(\Lambda,\Lambda') \nn\\ 
&&+\frac{\lambda}{5!} \int d\Lambda_A ..\Lambda_E \sum_{J_i,A_i} \vphi^{J_1..J_4}_{A_1..A_4}(\Lambda_A)...\vphi^{J_{10}..J_1}_{A_{10}..A_1}(\Lambda_E) \prod_i D^{J_i}_{00}(\Lambda_{s(i)}\Lambda_{t(i)}^{-1}) \\ &=& \sum_{J_i,A_i}
\vphi^{J_i,J}_{A_i,B_i,k,l}\vphi^{J_i,J'}_{A_i,C_i,m,n}\left(\int d\Lambda d\Lambda' D^{J}_{kl}(\Lambda)D^{J'}_{mn}(\Lambda')\f{1}{\Delta_{J_i}} {\cal C}_{B_i C_i}^{J_i}(\Lambda,\Lambda')\right) \nn\\ &&+\frac{\lambda}{5!}  \sum_{J_i,A_i} \vphi^{J_1..J_4,J_A}_{A_1..A_4,m_A,n_A}...\vphi^{J_{10}..J_1,J_E}_{A_{10}..A_1,m_E,n_E} \left(\int d\Lambda_A ..\Lambda_E D^{J_A}_{m_A n_A}(\Lambda_A)...D^{J_E}_{m_E n_E}(\Lambda_E) \prod_i D^{J_i}_{00}(\Lambda_{s(i)}\Lambda_{t(i)}^{-1})\right)\nn,
\ees
in momentum space, from which it is immediate to read out the vertex amplitude for our Feynman diagrams encoding the interaction of our field $\vphi$, given in configuration space by:
\be
\mathcal{I}_v (g_{\mid v}, \tilde{g}_{f\mid v}, \Lambda_{e \mid v}) = \prod_{f\mid v} \delta(g_{f\mid v}\Lambda_{e1}^{-1}\Lambda_{e2}\tilde{g}_{f\mid v}^{-1}) \label{eq:vertex}
\ee

while the propagator is given by the expression: 

\be
\pp(g_i,G; \tilde{g}_i, G') = \int d\Lambda\int d\Lambda' \prod_i \delta(g_i \Lambda^{-1},\tilde{g}_i\Lambda'^{-1}) \, f^{-1}(G\Lambda^{-1},G'\Lambda'^{-1})
\ee 
in configuration space, and by:
\be
\pp(J_i,A_i, B_i,J,mn; \tilde{J}_i,\tilde{A}_i, C_i,J',kl)=\prod_i\delta_{J_i,\tilde{J}_i}\prod_i \delta_{A_i,\tilde{A}_i} \int dGdG' D^{J}_{mn}(G)D^{J'}_{kl}(G') \pp^{J_i}_{B_i C_i}(G,G')
\ee
in momentum space, with $\pp^{J_i}$ having been defined above. This is for the general case; for the specific choice of coupling function proposed above, i.e. the heat kernel on the group, the analogous expressions are easily written down: the interaction term is unaltered, while the propagator takes the form:
\be
\pp^\beta(g_i,G; \tilde{g}_i, G') = \int d\Lambda\int d\Lambda' \prod_i \delta(g_i \Lambda^{-1},\tilde{g}_i\Lambda'^{-1}) \, \kk^{\beta}(G\Lambda^{-1},G'\Lambda'^{-1})
\ee
that reduces as we had discussed to the usual Barrett-Crane propagator for $\beta\rightarrow \infty$ and gives instead a BF-type product of five deltas if the limit $\beta\rightarrow 0$ is taken instead. 

Analogously, one can impose the simplicity projectors in the kinetic term and work fully in the homogeneous space; the construction is not altered substantially, but checking the two limiting properties in $\beta$ for the resulting Feynman diagrams is much less straightforward, especially in configuration space.   

The partition function is then defined in terms of its Feynman expansion, as usual, and it is given by:
\be
Z(\lambda,\beta) = \int \mathcal{D}\vphi\; e^{- S[\vphi]} = \sum_{\Gamma} \frac{\lambda^N}{sym(\Gamma)} \,Z_{\Gamma}[\beta]
\ee
where $N$ is the number of vertices in the graph and $sym$ its order of automorphisms, and with the amplitude for each graph being given by the appropriate convolution of vertex amplitudes and propagators as:

\be
Z_{\Gamma}[\beta]= \left(\prod_e \prod_{f\mid e} \int dg_{f\mid e} \int d\tilde{g}_{f\mid e}\right) \left( \prod_e \int dG_e \int dG'_e\right) \prod_e \pp^{\beta}(g_{f\mid e},\tilde{g}_{f\mid e}, G_e,G'_e) \prod_v \mathcal{I}_v(g_{f\mid v}, \tilde{g}_{f\mid v}, G_{e\mid v})
\ee
 
It is now easier to understand the structure of the resulting models; in particular, we see how indeed, in the $\beta\rightarrow 0$ limit of the coupled model, the fifth variables of the fields, having the interpretation of normals to the tetrahedra in the reference frames defined by each 4-simplex, are forced to be mapped trivially from one 4-simplex to the next. It is indeed a flat model, with a rigid coupling between 4-simplices and similar to a BF theory in that allows {\it only} such a flat configuration at least for what concerns the extra variables. It is not however a GFT formulation of a true BF theory for a 5-dimensional spacetime, and differs of course also from a true BF theory in 4-dimensions, in that: 1) the first four variables of the field are projected down to the homogeneous space $\SO(3,1)/\SO(3)$ while maintaining the gauge invariance under the full Lorentz group; 2) the fifth variables in the interaction term, i.e. in each 4-simplex, are completely decoupled.

It is important to stress that not only the new model gives a simple way of interpolating between the Barrett-Crane model and the flat model by moving in parameter space, i.e. by following a renormalisation group flow, but most importantly perhaps it allows for perturbation analysis around these limiting configurations, i.e. it permits to study the physics of the model near the \lq decoupled phase', i.e. the ordinary Barrett-Crane model, and near the \lq strongly coupled phase' instead. Interestingly, one would expect to see propagating \lq gravitons', i.e. propagating perturbations in the curvature, in the almost flat case, thus in the strongly coupled phase, and \lq confined gravitons' in the decoupled phase, when perturbing around the usual Barrett-Crane configurations. In the almost flat case it would correspond to a definition of a gravity theory in terms of a perturbative expansion around a flat connection configuration; in both cases it would amount to a fully covariant perturbation expansion, analogous to the one proposed in \cite{laurent-artem}, and indeed the exact relationship with the model in \cite{laurent-artem} should be analysed and we leave it for further study. In a renormalisation group interpretation, according to which the $\beta$-dependent model would possess two physically very different phases, the Barrett-Crane or decoupled phase and the strongly coupled BF-type one, this perturbation expansion would then be an expansion around the two possible fixed points of the renormalisation group flow. 
Moreover, we are going to show in the following that the $\beta$-dependent model bases on the heat kernel coupling lend itself to a further generalisation which could be the basis for further progress concerning the renormalisation of spin foam models. But before discussing this further generalised model, we would like to show how the five argument formalism for the field theory described above and the associated mathematical structures arise naturally when particles are inserted in the spin foam model.

\subsection{About particle insertions in the Barrett-Crane model} \label{sec:partBC}
In flat spacetime, a basis of one-particle states is given by momentum
eigenstates $\psi_{p,\sigma}$, where $p$ is the four-momentum and
$\sigma$ the other degrees of freedom - the spin. A Lorentz
transformation $U(\Lambda)$ will map a state of momentum $p$ to a
state of momentum $\Lambda p$. Then we can choose a reference momentum
$k^\mu$ and define the states $\psi_{p,\sigma}$ from the action of Lorentz
boosts on the reference state:

$$
\psi_{p,\sigma}=N(p)U(s(p))\psi_{k,\sigma},
$$
where $N(p)$ is a normalization factor and $s(p)$ is a Lorentz
transformation mapping $k$ to $p$.
The usual normalization is $N(p)=\sqrt{k_0/p_0}$ so that the scalar product
is normalized
$\la\psi_{p,\sigma},\psi_{p,\sigma}\ra=
\delta_{\sigma\sigma'}\delta^{(2)}(\vec{p}-\vec{p'})$.
For massive particles, we usually take $k=(1,0,0,0)$ and the little
group of Lorentz transformations $W$ leaving $k$ invariant is the
rotation group $\SU(2)$. Then the action of a Lorentz transformation
reads:
$$
U(\Lambda)\psi_{p,\sigma}=N(p)U(s(\Lambda p))U(W(\Lambda,p))\psi_{k,\sigma},
$$
with $W(\Lambda,p)=s(\Lambda p)^{-1}\Lambda s(p)$ living in the little group.
From this, we see that it is enough to postulate the action of the little group:
$$
U(W)\psi_{k,\sigma}=\sum_{\sigma'}D_{\sigma',\sigma}(W)\psi_{k,\sigma'},
$$
in order to get the action of an arbitrary Lorentz transformations:
$$
U(\Lambda)\psi_{p,\sigma}=
\left(\f{N(p)}{N(\Lambda p)}\right)
\sum_{\sigma'}D_{\sigma',\sigma}(W(\Lambda,p))\psi_{\Lambda p,\sigma'}.
$$
The {\it spin} $s$ is the choice of a irreducible representation of the little group $\SU(2)$,
which defines the coefficient $D_{\sigma',\sigma}$, and $\sigma$ is the angular momentum. Then we have induced a representation of the Lorentz group from a $\SU(2)$ representation.

\medskip

To include particles in the Barrett-Crane model, we first look at the
particle insertion on a 4-simplex. The one-4-simplex amplitude is
simply the evaluation of its boundary spin network. Thus we need to
include the particles in a spin network wave function. Inserting
particles at all vertices of the spin network graph, the wave function
is then a function of $E$ holonomies $U_e\in\SL(2,\C)$, describing the
parallel transport between vertices and carrying the gravitational
degrees of freedom, and $V$ group elements
$\Lambda_v\in\SL(2,\C)/\SU(2)$ in the hyperboloid $\hh$, defining the
state of the particles, where $E$ and $V$ are respectively the number
of edges and vertices of the graph. Considering the transformations of
holonomies and momenta under gauge transformations, we are looking at
the following gauge invariant functions:
\be
\psi(U_1,..,U_E,\Lambda_1,..,\Lambda_V)=
\psi(g_{s(1)}^{-1}U_1g_{t(1)},..,g_{s(E)}^{-1}U_Eg_{t(E)},g_1\Lambda_1,..,g_V\Lambda_V),
\qquad
\forall g_v\in(\SL(2,\C))^{\otimes V}.
\ee
Such wave functions have already been considered in the context of
(covariant) loop quantum gravity, and in \cite{proj} a basis for
the Hilbert space of such wave functions ($L^2$ invariant functions
for the Haar measure on the Lorentz group) is defined in terms of {\it projected spin
  networks}. Roughly, these are spin networks labeled with $\SL(2,\C)$
representations on every edge, and with $\SU(2)$ representations on
every edge and $\SU(2)$ intertwiners at every vertex. In the case that
we choose the $\SU(2)$ to be all the trivial scalar representation, we
find back the simple spin networks of the standard Barrett-Crane model
\cite{proj,us}:
$$
\psi_{J_e}(U_1,..,U_E,\Lambda_1,..,\Lambda_V)=\prod_e K^{J_e}(\Lambda_{s(e)}^{-1}U_e\Lambda_{t(e)}),
$$
where we note $s(e)$ and $t(e)$ the source and target vertices of the edge $e$.
To each vertex of the simple spin network is associated the
$\Lambda_v$ dependent $\SU(2)$ intertwiner between the $\SL(2,\C)$
representations $R^{J_i}$:
\be
\bigotimes_i \la J_i,\Lambda_v,j=m=0|\,:\bigotimes_i R^{J_i}\,\arr\,\C,
\label{simple}
\ee
where $i$ runs over all the edges meeting at the vertex
$v$. $|J_i,\Lambda,j=m=0\ra=D^{J_i}(\Lambda)W_{J_i}$ is the vector in
$R_{J_i}$ which is invariant under the $\SU(2)$ subgroup of
$\SL(2,\C)$ leaving invariant the direction $\Lambda\in\hh$.

We recognize in the above structure the basis of states in which we decompose the 4-valent field of the usual group field theory formulation of the Barrett-Crane model, or, equivalently, the basis states arising in the trivial 5-valent extension of the same group field theory, discussed at the beginning. The field is indeed decomposed in 4-valent simple intertwiners of the Lorentz group, i.e. carrying simple representations of the Lorentz group on every link, and trivial $SU(2)$ representations on the same links, and a $\Lambda$ dependent $SU(2)$ intertwiner at each vertex, with the various $\Lambda$'s being integrated over at the end when evaluating the spin network amplitude. 

There is another way of building a basis of the space of such
invariant wave functions, which makes the link with the particle
insertions clearer. In fact the construction is very close to the way spinning particles are coupled in 3d quantum gravity (see \cite{pr1,pr3,3d} for the construction of coupled spin foam amplitudes, and \cite{particle, dj} for the group field theory formalism). To each $n$-valent
vertex of the spin network graph, we associate a $(n+1)$ valent intertwiner $\ii$
which intertwines the $n$ edges meeting at the vertex and an extra
edge describing the particle insertion. The particle is then
characterized by a choice of a Lorentz representation $J$ and a vector
$\vv^J$ in the corresponding Hilbert space $R^J$. At each vertex, the
intertwiner $\ii_v$ is a $\SL(2,\C)$ invariant tensor in $\otimes_i
R^{J_i}\,\otimes R^J$. The wave functions is then the evaluation of
the holonomies on the spin network with the following tensors at each
vertex:
\be
\ii_v(\,\cdot\, ,D^J(\Lambda)\vv^J)
\,:\bigotimes_i R^{J_i}\,\arr\,\C.
\label{particle}
\ee
To describe a particle of spin $s$, we simply choose the vector
$\vv^J$ to lie in the subspace of $R^{J}$ corresponding to the
$\SU(2)$ representation of spin $j\equiv s$. Therefore, to describe a
spinless particle, we choose the $\SU(2)$ invariant vector
$\vv^J\equiv W^J$.

We recognize in this structure the basis of states arising in the mode expansion of our 5-valent field, in the non-trivial extension of the group field theory presented above.


These two basis, projected spin networks and particle insertions,
correspond thus to two choices of Fourier modes for the decomposition of
the $\phi(g_i,G)$ in our group field theory context. It is easy to see
that the simple intertwiner \Ref{simple} corresponds to the sum over
all possible particle insertions \Ref{particle} for fixed
representations $J_i$, i.e. sum over all possible $J$ and
$(n+1)$-valent intertwiners.
Indeed:
\be
\bigotimes_i |J_i,\Lambda, j=m=0\ra
=\prod_i D^{J_i}_{A_iB_i}(\Lambda)W^{J_i}_{B_i}
=\sum_J\sum_{\ii}\ii^{J_1..J_n,J}_{A_1..A_n,A}D^J_{AB}(\Lambda)W^J_B,
\ee
where we are summing over an orthonormal basis of $(n+1)$ valent intertwiners between the representations $J_1,..,J_n,J$.
Nevertheless, more generally, if we want to include a particle of some
given fixed spin $s$, we will have to go beyond simple spin networks
and allow arbitrary $\SU(2)$ representations in the projected spin
network basis. Here we will not write in details the explicit change
of basis between the two choices of modes. 
Instead we would like to
stress again that the generalized group field theory introduced in the
previous section with an extra group variable for each field (or
quantized tetrahedron) would allow to compute transition amplitudes
for gravity plus particles in the spin foam formalism.
This also means that, on the one hand, as we anticipated, this kind of mathematical structures would probably play a role in any group field theoretic formulation of 4d quantum gravity coupled to point particles, and, on the other hand, that the insertion of point particle in the Barrett-Crane model, or a similar one, would provide the model with a non-trivial coupling between spacetime atoms, i.e. between 4-simplices, possibly of the type we have presented here. 

\section{A new type of group field theory: turning the coupling into a variable} \label{sec:gengft}
We have just seen that the choice of the heat kernel
$\kk^{\beta}(g,g')$ for the propagator term, relating the two normals
corresponding to the same tetrahedron in two neighboring 4-simplices,
generates a model that interpolates nicely between the usual
Barrett-Crane model and a strongly coupled model where the only
geometric configuration allowed is flat space, with all the normals
being identified. This very same choice leads naturally to a
further generalization of the group field theory and of spin foam
models, that we believe has even more interesting properties and may
be a good starting point for further investigations.

Consider any given spin foam (i.e. Feynman graph) amplitude for the
above model, with the heat kernel with parameter $\beta$ on each dual
edge connecting the two group variables corresponding to the fifth
variables of the field, i.e. to the tetrahedron normals. Now
re-interpret the $\beta$ as a {\it variable} of the model, as opposed
to a freely specified {\it parameter}. This implies that there is an
integral over it, with range from zero to infinity in the definition
of the amplitudes; for the same reason, the two limits mentioned for
this interpolating model correspond now to two possible approximations
one can make for this integral, considering only the region of
integration near zero or the asymptotic region at infinity; this last
one describes perturbations around the Barrett-Crane configurations,
while the first corresponds to a strong coupling region. 

Even more interesting is the group field theory derivation of this
extended model, with $\beta$ treated as a variable. First of all
notice that, if $\beta$ has to be considered as a variable, the
definition of the field has to be extended to a function 
$$ \phi(g_1,...,g_4;G,\beta) : SO(3,1)^5\times \mathbb{R} \rightarrow
\mathbb{C}, $$ with the global invariance under $SO(3,1)$
transformations and the invariance under $SO(3)$ transformations of
the first four arguments to be then imposed in the action. 

Now, considering just the Feynman amplitudes in configuration space, i.e. not performing any mode expansion, we have now a propagator of the form:
\be
\pp(g_i,g'_i,G,G',\beta)=\left(\prod_i\delta(g_i,g'_i)\right) \kk(G,G';\beta),
\ee
where we switched to a different notation for the heat kernel to
highlight the fact that now $\beta$ has the interpretation of a \lq\lq
Euclidean time \rq\rq for a particle living on the homogeneous space
to which $G$, and $G'$ belong, and the heat kernel function indeed
propagates it from the point $G'$ at time 0 to the point $G$ at (Euclidean) time
$\beta$. We assume that $\kk(G,G';\beta)$ is zero for $\beta<0$, in order to have a well-defined mode expansion for the propagator, so we effectively work with $\kk^+(G,G';\beta)=\theta(\beta)\kk(G,G';\beta)$, where $\theta$ is the Heaviside function. 

It is easy to identify the kinetic term from which this
propagator comes. Indeed the heat kernel with \lq\lq time\rq\rq variable $\beta$ is
simply the propagator for the heat equation 
\be
\left( -\frac{\partial}{\partial \beta} + \nabla \right) \psi(g,\beta) = 0,
\ee
with $\nabla$ being the Laplace-Beltrami operator on the group manifold.
Therefore the appropriate kinetic term of the group field theory action is identified with:

\be
S_{kin}[\phi] = \prod_i \int dg_i \int dG\int d\beta \;\phi(g_1,...,g_4;G,\beta)\left[\left( -\partial_\beta + \nabla_G \right) \phi\right](g_1,...,g_4;G,\beta).
\ee
and the kinetic operator with:
\be
K(g_i,G,\beta; \tilde{g}_i,G',\beta')= \left( \prod_i \delta(g_i,\tilde{g}_i) \right) \left( -\partial_\beta + \nabla_G\right)\delta(G,G')\delta(\beta,\beta').
\ee

In the following we will actually use a symmetrized version of the propagator above, so to erase any asymmetry between the two possible sign of the $\beta$ variable entering in it, so we will work with a propagator of the form:
\be
\pp_H(g_i,g'_i,G,G',\beta)=\left(\prod_i\delta(g_i,g'_i)\right) \left[ \theta(\beta)\kk(G,G';\beta) + \theta(-\beta) \kk(G,G',-\beta)\right].
\ee
This has an interpretation in terms of different transition amplitudes one can define for quantum gravity in a group field theory context \cite{generalised}.

We can now consider the interaction term for our generalized group field theory. A first choice is to define the dependence on the extra variable $\beta$ in each field to be trivialized, i.e. a different $\beta$ enters in the various fields in the interaction term, but no relation is assumed between them; the interaction is then of the form:

\bes
\lefteqn{S_{int}^{(1)}[\phi]=\int d\beta_A ...d\beta_E\,\int d\Lambda_A ...d\Lambda_E \int \prod_{i=1}^{10} dg_i\prod_i \int dh_i d\tilde{h}_i\,
\phi(g_1 h_1\Lambda_A,g_2 h_2\Lambda_A,g_3 h_3\Lambda_A,g_4 h_4\Lambda_A,\Lambda_A, \beta_A)} \nonumber \\ &&\phi(g_4 \tilde{h}_4 \Lambda_B,g_5 h_5\Lambda_B,g_6 h_6 \Lambda_B,g_7 h_7 \Lambda_B,\Lambda_B, \beta_B)\; \phi(g_7\tilde{h}_7\Lambda_C,g_3 \tilde{h}_3\Lambda_C,g_8 h_8 \Lambda_C,g_9 h_9 \Lambda_C,\Lambda_C, \beta_C) \nonumber
\\ &&\phi(g_9 \tilde{h}_9 \Lambda_D,g_6\tilde{h}_6\Lambda_D,g_2\tilde{h}_2\Lambda_D,g_{10}h_{10}\Lambda_D,\Lambda_D, \beta_D)\; \phi(g_{10}\tilde{h}_{10}\Lambda_E,g_8\tilde{h}_8\Lambda_E,g_5\tilde{h}_5\Lambda_E,g_1\tilde{h}_1\Lambda_E,\Lambda_E,\beta_E).
\ees

This is an admissible choice for the interaction term; however, the resulting dependent on the $\beta$ variables is somehow unsatisfactory: there are two $\beta$ variables associated with each tetrahedron in the dual formulation of the associated Feynman graphs, i.e. two $\beta$ variables for each edge of the Feynman graph; the amplitudes depend only on the difference between them, but this dependence is \lq\lq localised\rq\rq along the edge of the Feynman graph only, i.e. the variables appearing there do not appear anywhere else in the amplitude, so the integral can actually be performed separately in each edge, leading to a modified but $\beta$-independent propagator. Explicitly, this modified $\beta$-independent propagator of this model is given by:
\be
\pp_H(g_i,\tilde{g}_i,G,G')= \prod_i \delta(g_i,\tilde{g}_i)\,\int d\beta \int d\tilde{\beta}\;\; \pp_H(G,G';\beta-\tilde{\beta}), 
\ee
where $\beta$ and $\tilde{\beta}$ are the two \lq\lq Euclidean time\rq\rq variables associated to the give edge. The integral can be performed easily if one uses the mode expansion of the heat kernel, assuming that one can exchange the sum over representation and the integral over the $\beta$ variables, leading to:
\bes
\lefteqn{\pp_H(g_i,\tilde{g}_i,G,G')=\,2\prod_i \delta(g_i,\tilde{g}_i)\,\int_0^\infty d(\beta -\tilde{\beta}) \;\;\kk(G,G';\beta-\tilde{\beta})=} \nonumber \\ &=& 2\prod_i \delta(g_i,\tilde{g}_i)\,\int_0^\infty d(\beta -\tilde{\beta}) \sum_J \Delta_J e^{-(\beta - \tilde{\beta}) C_J} \chi^{J}(G G'^{-1}) = 2\prod_i \delta(g_i,\tilde{g}_i)\,\sum_J \frac{\Delta_J}{C_J}\, \chi^{J}(G G'^{-1}),
\ees  
where $C_J$ is the non-zero quadratic Casimir for simple representations of the Lorentz group
(with an appropriate analytic continuation needed to regularise the integral), to which the sum refers, $\chi^J$ is the character of the representation, and we have discarded an infinite factor coming from the integral over $\beta + \tilde{\beta}$ and that is clearly pure gauge, the amplitude being completely independent of such a variable.

One can still consider the different regimes of integration over $\beta - \tilde{\beta}$, i.e. the regime in which the difference is large and the model then approximates the usual Barrett-Crane model, and the one in which the difference is very small, and the model gives a BF-type behavior; however, such a localized dependence on the variables and the triviality of the amplitudes with respect to this dependence, to the point that the $\beta$ variables can be quite simply eliminated from the amplitudes by integrating them out, make this choice of interaction term somewhat unpleasant, and we are led to look for a different and less trivial choice.

If we want a relationship between the various $\beta$ variables in the fields in the same vertex term, then the simplest choice is to choose them to be all equal; the interaction term in the action would then be:

\bes
\lefteqn{S_{int}^{(2)}[\phi]=\int d\beta \; \int d\Lambda_A ...d\Lambda_E \int \prod_{i=1}^{10} dg_i\prod_i \int dh_i d\tilde{h}_i\,
\phi(g_1 h_1\Lambda_A,g_2 h_2\Lambda_A,g_3 h_3\Lambda_A,g_4 h_4\Lambda_A,\Lambda_A, \beta)} \nonumber \\ &&\phi(g_4 \tilde{h}_4 \Lambda_B,g_5 h_5\Lambda_B,g_6 h_6 \Lambda_B,g_7 h_7 \Lambda_B,\Lambda_B, \beta)\; \phi(g_7\tilde{h}_7\Lambda_C,g_3 \tilde{h}_3\Lambda_C,g_8 h_8 \Lambda_C,g_9 h_9 \Lambda_C,\Lambda_C,\beta) \nonumber
\\ &&\phi(g_9 \tilde{h}_9 \Lambda_D,g_6\tilde{h}_6\Lambda_D,g_2\tilde{h}_2\Lambda_D,g_{10}h_{10}\Lambda_D,\Lambda_D, \beta)\; \phi(g_{10}\tilde{h}_{10}\Lambda_E,g_8\tilde{h}_8\Lambda_E,g_5\tilde{h}_5\Lambda_E,g_1\tilde{h}_1\Lambda_E,\Lambda_E,\beta)
\ees     

This choice is more satisfactory for several reasons; first of all it avoids the problems mentioned above for $S_{int}^{(1)}$, in that the dependence on $\beta$ is much less trivial and not localized along each edge; second, with the interpretation of $\beta$ as a interaction coupling parameter in each spin foam amplitude, this choice corresponds to having a different coupling parameter in each spacetime building block, i.e. 4-simplex, with the amplitude for the interaction between these building blocks depending only on the difference between the parameters associated to them; third, if one keeps in mind instead the interpretation of the $\beta$ variable as a kind of \lq\lq Euclidean time\rq\rq, then the choice of $S_{int}^{(2)}$ made above is actually the most sensible one, as it corresponds to an interaction between the five fields corresponding to the five tetrahedra in 4-simplex that is {\it local} in this time variable.     
Note that also in this case we have a different coupling of spacetime atoms, i.e. 4-simplices, for each edge of the spin foam, but with a non-trivial relation among those referring to the same 4-simplex.
The vertex amplitude in configuration space corresponding to the interaction term $S_{int}^{(2)}[\phi]$ is therefore still given by \Ref{eq:vertex}:
\be
\mathcal{I}_v(g_f,\tilde{g}_f, \Lambda_e, \beta)= \prod_{f\mid v} \delta(g_{f\mid v}\Lambda_{e1}^{-1}\Lambda_{e2}\tilde{g}_{f\mid v}^{-1}),
\ee
where the dependence on $\beta$ is limited to the fact that the same $\beta$ should appear in any mode expansion of the fields with respect to this variable, that we do not perform explicitly, and in that the same $\beta$ has to appear in any propagator connecting this vertex to any other one in the Feynman expansion. This is given of course by:

\be
Z(\lambda) = \int \mathcal{D}\phi\; e^{- S[\phi]} = \int \mathcal{D}\phi\; e^{- S_{kin}[\phi] - \frac{\lambda}{5!} S_{int}^{(2)}[\phi]} = \sum_{\Gamma} \frac{\lambda^N}{sym(\Gamma)} \,Z_{\Gamma}, \label{eq:Zvar}
\ee
with:
\be
Z_{\Gamma}= \left(\prod_e \prod_{f\mid e} \int dg_{f\mid e} \int d\tilde{g}_{f\mid e}\right) \left( \prod_e \int dG_e \int dG'_e\right) \prod_v \int d\beta_v \prod_e \pp_H(g_{f\mid e},\tilde{g}_{f\mid e}, G_e,G'_e, \beta_{e}, \tilde{\beta}_e) \prod_v \mathcal{I}_v(g_{f\mid v}, \tilde{g}_{f\mid v}, G_{e\mid v}, \beta_v).
\ee

We think this new type of group field theory model possesses many interesting features, and the resulting spin foam amplitudes should be analysed in more detail, but we leave this analysis for future work; now we prefer to simply highlight some of its general aspects, its possible interpretation and uses. Note for example that in this new model the coupling relative to different edges of the spin foam, i.e. too different tetrahedra, may be different so that it is possible to study models that are almost flat, but not quite, and with small {\it different} curvatures in different regions of space.

First of all the model presented above can be seen at the same time as an extension and as a special case of the new type of group field theories studied and to be presented in \cite{generalised}; there a \lq\lq proper time\rq\rq parametrisation of group field theories is studied in order to achieve a generalized formulation of them from which both orientation dependent and orientation independent spin foam models \cite{us,feynman} can be obtained; the field is defined to be of the general form:
\be
\vphi(g_1, s_1; g_2, s_2; g_3, s_3; g_4, s_4) : \left( \SO(3,1) \times \mathbb{R}\right)^{\times 4}
\ee  
and on this field both a simplicity projector onto $\SO(3,1)/\SO(3)$ for the four arguments and a global \lq\lq gauge invariance\rq\rq projector, under $\SO(3,1)\times \mathbb{R}$ transformations acting simultaneously (diagonally) on the four arguments, are applied at the level of the action; with respect to a model based on this type of field, the one we have outlined above represents a restriction, because the extension of the domain of the first four arguments of the field from $\SO(3,1)$ to $\SO(3,1)\times \mathbb{R}$, with an extra \lq\lq proper time\rq\rq variable, is dropped; but it is also a generalization, in the same sense as the model presented in section \Ref{sec:genBC} is a generalization of the Barrett-Crane model, because it extends the field to a five argument one, again by basically turning the variables enforcing the gauge invariance of the four argument field into true extra variables of it, and extending appropriately the gauge invariance to be imposed.   
Indeed we expect a direct generalisation along these lines of the type of models studied in \cite{generalised} to be straightforward and to lead to a model of the same type as the one defined by the partition function \Ref{eq:Zvar}.  The main difference is however that what appears in the model defined by \Ref{eq:Zvar} is a kind of \lq\lq Euclidean time\rq\rq, not a Minkowskian one, so that the corresponding differential operator appearing in the action is of heat- or diffusion-type, and not of Schr\"oedinger type; this is needed for maintaining the property that the model reduces to the Barrett-Crane one in the large $\beta-\tilde{\beta}$ limit. 

It is interesting at this point to discuss briefly the classical equations of motion for the field $\phi$  coming from the action $S[\phi] = S_{kin}[\phi] + S_{int}^{(2)}[\phi]$. These are:
\bes
\lefteqn{\left(-\partial_\beta + \nabla_G\right)\phi(g_1,g_2,g_3,g_4,G,\beta) +} \nonumber \\  &+&\frac{\lambda}{5!} \int dg_5...dg_{10}\;\vphi(g_1,g_5,g_6,g_7,G,\beta)\phi(g_2,g_5,g_8,g_9,G,\beta)\vphi(g_3,g_6,g_8,g_{10},G,\beta)\vphi(g_4,g_7,g_9,g_{10},G,\beta)= 0,
\ees 
which is indeed the form of a diffusion or heat transfer equation in Euclidean time $\beta$, with an additional time dependent \lq\lq driving potential\rq\rq, for what concerns the fifth extra arguments; the dynamics defined by this equation for the other arguments of the field is basically the usual one, modulo the modified gauge invariance requirement.  However, the physical interpretation of such an equation is not straightforward from a quantum gravity perspective, mainly because, while the field $\phi$ itself has the (quantum) geometric interpretation of a (second) quantized tetrahedron, the geometric interpretation of $\beta$ is unclear at the present stage. On the other hand, recalling the interpretation of $\beta$ as a coupling parameter, the above equation can be interpreted as a kind of renormalisation group equation for our coupled model and it is quite striking, in our opinion, that what is a renormalisation group equation from the point of view of the spin foam amplitudes as presented in the previous section is just an equation of motion from the point of view of the generalised group field theory described above. Indeed, this new type of group field theory can be a useful starting point, we think, for tackling the issue of renormalisation of spin foam models, or, better, of the group field theories generating them as Feynman graphs and amplitudes. Therefore we conclude now by discussing how this issue can be tackled from the starting point provided in this paper.

\section{Outlook: renormalisation of spin foam models} \label{sec: renorm}

The final goal is to understand the semi-classical behavior of the spin foam quantum gravity, recover the standard effective physics on flat or curved space-time and compute the corrections due to the quantum fluctuations of the gravitational field. This is a long-term program, and a first step in this direction would be to understand the coarse-graining and renormalisation of the spin foam models such as the Barrett-Crane model. We could expect a semi-classical space-time to be built out of a large number of elementary Planck-size blocks (the 4-simplices) and the macroscopic effective spin foam model to have a structure similar to that of the microscopic theory, but with a slightly different form of the amplitudes and a different value of the couplings: the Barrett-Crane model would then be only be a particular model along the renormalisation flow. We propose to use our generalized Barrett-Crane with arbitrary couplings between 4-simplices as an ansatz to study explicitly this renormalisation flow.
 
\subsection{About the renormalisation of the coupling}

The Barrett-Crane amplitude for a 4-simplex is a function of the 10 representation labels $\rho_f\in\R_+$ living on its faces and of 5 variables $\Lambda_T\in\SO(3,1)/\SO(3)$ attached to each tetrahedron. We note it $\aa(\rho_f,\Lambda_T)$. The $\rho_f$'s have the interpretation of the area of the corresponding triangle while the $\Lambda_T$'s are thought the (time-like) normal to the corresponding tetrahedron. Then the Barrett-Crane $\{10\rho\}$ symbol can be considered as a (quite peculiar) quantum amplitude for a single quantum 4-simplex described in terms of 1st order Regge calculus\cite{us,barrett,barrettruth}. A full spin foam is constructed by gluing 4-simplices together. As explained above, the representation label attached to a triangle will be the same for all 4-simplices who share that triangle. However, the normal variables $\Lambda_T$ are attached to the corresponding tetrahedron in the reference frame of a given 4-simplex: they are not shared between 4-simplices and we will not denote them $\Lambda_{T,\sigma}$ referring to both the tetrahedron $T$ and the 4-simplex $\sigma$ to which it belongs. Then the Barrett-Crane spin foam amplitude is the product of the 4-simplex amplitudes summed over all assignments $\{\rho_f,\Lambda_{T,\sigma}\}$.

Let us consider a particular tetrahedron $T_0$ in a fixed spin foam. It belongs to two 4-simplices, $\sigma_1$ and $\sigma_2$. The two normal variables for $T_0$, $\Lambda_{T_0,\sigma_1}$ and $\Lambda_{T_0,\sigma_2}$, are  a priori independent. They are indeed both integrated out independently to get the Barrett-Crane amplitude. Nevertheless, if we compute the spin foam amplitude summing and integrating over all representation labels $\rho_f$ and all other normal variables $\Lambda_T$, $T\ne T_0$, we will obtain a probability amplitude on  the space of these two variables, $\Lambda_1$ and $\Lambda_2$ in short; this amplitude will not be factorised in general, which implies that they will not be treated as fully independent by the model after this sort of \lq\lq renormalisation procedure\rq\rq involving the mentioned summation over representations and normal variables. Therefore, with the aim of understanding the renormalisation of the Barrett-Crane model, it is natural to work in a generalized formulation allowing for generic coupling between the two normals associated to each tetrahedron. This should be intended as an effective way for encoding the coupling between these variables that would result from renormalisation, before the actual renormalisation procedure is carried out. 
Another way to state the same point is the following. The Lorentz transformation mapping $\Lambda_{T_0,\sigma_1}$ to $\Lambda_{T_0,\sigma_2}$ can naturally be interpreted as the parallel transport between the two reference frames associated to the two 4-simplices $\sigma_1$ and $\sigma_2$. In the "fundamental" model, this parallel transport is arbitrary, and therefore $\Lambda_1$ and $\Lambda_2$ are integrated over independently. Nevertheless, under renormalisation of that fundamental model, we expect that the effective probability amplitude for this parallel transport evolves to a non-trivial correlation $\cc(\Lambda_1,\Lambda_2)$ between $\Lambda_1$ and $\Lambda_2$.

\medskip

From this perspective, it would be interesting to look for the existence of a fixed point of the renormalisation of the generalized coupled Barrett-Crane model, which we described above. We think that the group field theory provides us with the right framework to address this issue. A first result would be to check that the uncoupled spin foam model is not a fixed point (or at least not a stable fixed point) and then whether or not the rigidly coupled model is a fixed point. This rigid coupling, or strong coupling limit, $\cc(\Lambda_1,\Lambda_2)\sim \delta (\Lambda_1^{-1}\Lambda_2)$ would corresponds to a flat model, as we said: the 4-simplices would be glued with a trivial parallel transport, so that the spin foam should be describing a flat space-time \cite{us}.

The recent study \cite{wade} of the Lorentzian Barrett-Crane partition function in terms of the normal variables $\Lambda$'s should be particularly relevant to this project: the author looks at the properties of the amplitudes obtained by integrating first the representation labels $\rho_f$ and uses them to present a new proof of the finiteness theorems for the Barrett-Crane amplitudes. It would actually be rather natural to include arbitrary couplings between the $\Lambda$'s in such a reformulation of the model.

We also proposed a more specific group field theory for a restricted class of coupling given by the heat kernel. It should be simpler to study the coarse-graining properties of this model for which the couplings are described by a single real parameter. This model is close to a very interesting model proposed and studied by Oeckl \cite{robert}.
He introduces a new model replacing the $\delta$-functions constituting the edge and vertex amplitudes (in configuration space) in the spin foam quantization of the topological BF theory and the various Barrett-Crane-like models by heat kernels, and discusses the stability properties of the weak and strong coupling limits under coarse-graining moves generalizing the standard Pachner moves, i.e. under renormalisation. Our proposal is in the same spirit. However, while Oeckl's model corresponds to using the usual definition of the field and just changing the coupling between its arguments, the areas of triangles in momentum space, we first introduce a generalisation of the same field to five arguments, and then use explicitly the $\Lambda$ variables, representing the normal to the tetrahedra.
We feel that using the $\Lambda$ allows us to introduce the coupling between 4-simplices in a transparent way, easier to interpret physically. Moreover, Oeckl's model interpolates between the Barrett-Crane model and BF theory, while as we have discussed our $\beta\rightarrow 0$ limit does not give a true quantum BF theory but a similar theory in which the only configuration allowed for the (boost part of the) connection is the flat one. We believe it would be very interesting to analyse in more detail the similaritites and differences between these two modifications of the Barrett-Crane model, and in particular check whether the respective fixed points in the renormalisation flow in $\beta$ correspond to the same type of theory. 

The \lq\lq parametrised \rq\rq group field theory we have introduced last, in which the heat kernel coupling is treated as a variable of the theory, suggests (actually, forces) a different way of dealing with the same issues. In fact, as we said, the renormalisation flow of the coupling is encoded there in the {\it classical} equations of motion coming from the group field theory action; therefore, if we are to study the presence and nature of fixed points, we should check whether particular regimes for the coupling are favored by the classical equations of motion.

\medskip

We conclude this discussion by a speculation that the study of the renormalisation flow of the spin foam model could 
shed light on the issue of the diffeomorphism invariance of the spin foam amplitudes: diffeomorphisms may appear as renormalisation group transformations in the group field theory formulation since they seem to be related to Pachner moves in the state sum formulation of spin foams and to the Ward identities in the group field theory formulation. This was suggested in \cite{laurentgft,instantons}, following the results of \cite{laurentdiffeo}. Indeed Ward identities in quantum field theory relate the amplitudes of different Feynman diagrams. In the group field theory, they would then relate the amplitudes of different spin foams. It is likely that looking more precisely to this point would lead to a deeper understand of both the renormalisation of spin foams and the action of diffeomorphisms in this discrete quantum setting.  In the 2d spin foam case, the group field theory reduces to matrix models, for which the explicit link between the Ward identities of the path integral and the conformal symmetry is well-understood (see for example the review \cite{matrix}), so methods and ideas from matrix models can be of help in analysing this same issue for group field theories. Finally, it would be interesting to compare the resulting Ward identities relating different spin foam amplitudes to the background independent coarse graining scheme proposed by Markopoulou \cite{fotini}, using the Hopf algebra structure of 2-complexes in a way very similar to the Connes-Kreimer approach to the renormalisation of quantum field theory \cite{connes}.

\subsection{A numerical simulation project}

The coarse-graining of the coupling can well be studied numerically. First, we would need to look at the single 4-simplex amplitude $\aa(\rho_f,\Lambda_T)$. It would be interesting to look at the probability amplitude of the $\Lambda$ variables. We could restrict the study to regular 4-simplices where all the representation labels are identical, $\rho_f =\rho$. Fixing $\rho$ and integrating over three of the $\Lambda$ variables, we would get a probability amplitude on the remaining two $\Lambda$ variables, which would describe the correlations between two tetrahedra within the considered 4-simplex. Actually, due to the Lorentz gauge invariance of the 4-simplex, we would obtain a simple distribution on the angle between the two normals. A first issue is whether this distribution is peaked or not at a given angle. A second issue is whether this peak moves or not when rescaling the representation label $\rho$: this would show that the coupling would definitely be sensitive to the size of the 4-simplices (measured in Planck units). 

Still looking at a single 4-simplices, we could consider almost identical assignments of representation labels $\rho_f$. More precisely, singling out two tetrahedra, and keeping fixed the 7 representation labels attached to these tetrahedra, we could look at the evolution of the coupling between these tetrahedra when we change the 3 remaining representation labels. This would show how the geometry of a tetrahedron does not solely depend on the representation labels assigned to it but on the representation labels assigned to the 4-simplex to which it belongs: the intrinsic geometry of a tetrahedron is determined by the space-time geometry of the 4-simplex.

Let us point out that we could carry on this analysis unfolding the Barrett-Crane intertwiner at each tetrahedron in terms of $\SO(3,1)$ simple representations $\rho_{int}$ instead of the normal variable $\Lambda$. $\rho_{int}$ would be interpreted as describing the internal space geometry of the tetrahedron while the $\Lambda$ seem to describe its embedding in the surrounding spacetime. A first computation would be to study the correlations between the internal representation label $\rho_{int}$ of two tetrahedra within the same 4-simplex and see whether the maximal probability corresponds to small/large-small/large configurations. One might then hope that such correlations propagate within the spinfoam helped by the coupling between 4-simplices.

The next step would be to consider two 4-simplices or more and the coupling between two tetrahedra belonging to different 4-simplices. The coupling between the 4-simplices would then play a role, and therefore the new coupled model we have proposed in this paper can be of direct use. We could study how long-range correlations between spacetime points (tetrahedra) would depend on the coupling appearing between two glued 4-simplices. In this framework, we could either keep the representation labels fixed and look at the correlations induced solely by the coupling imposed by hand, or we could sum over internal representation labels (attached to the internal faces i.e not on the spin foam boundary) and we would look at the full effective correlations.

\section{Conclusions}
Let us summarize what we have achieved in this work. Our main result has been to construct a group field theory formalism that generalizes the Barrett-Crane model to allow a non-trivial coupling between the two normals to the tetrahedra in the two 4-simplices that share them; the way we have done it is simple: we have shown first how the usual Barrett-Crane model can be seen as the result of a quantization of a field theory on five copies of a group manifold, with no coupling among the fifth arguments of the field, and then we have modified this field theory to introduce a coupling. We have exhibited a specific choice of coupling function and therefore constructed a specific coupled model; this model interpolates nicely between the usual Barrett-Crane model, with the parallel transport between 4-simplices being completely randomized, and a flat or BF-like model, in which the only allowed connection is the flat one, according to the value that a new additional parameter of the theory, not present in the usual Barrett-Crane model, takes. The tuning of the coupling parameter therefore corresponds to a tuning of the degree of \lq\lq locality\rq\rq of the resulting spin foam amplitudes. This new coupled model is amenable for further study of correlations between simplices, both analytical and numerical, that can shed light on the presence or absence of local propagating degrees of freedom in spin foams, and can play an important role in studying renormalisation of the spin foam amplitudes under coarse graining. Moreover, it can be the basis for developing fully covariant perturbation expansions in the coupling parameter that do not make any use of background structures and therefore remain fully background independent from the quantum spacetime point of view. We have also discussed how the mathematical structures on which the coupled model is based are going to be relevant for studies of particle and matter insertions in the Barrett-Crane or similar spin foam models. Finally, we have gone further in this process of generalization, by showing how a new type of group field theory can be constructed in which the coupling parameter is promoted to a variable of the theory, on the same footing as group variables, leading to the presence of derivatives in the action and therefore to a different type of classical equations of motion for the field. We concluded by discussing the issue of renormalisation of spin foam models from the new perspective that the results of this work suggest, and how the new coupled model can be a useful new starting point for the application of renormalisation group techniques, and therefore for understanding the semi-classical properties of spin foam models.

\section*{Acknowledgements}
We would like to thank, for hospitality and beverages, Caffe' Nero in Cambridge and Cafe' 1842 in Waterloo, where part of this work was done. D.O. thanks also the Perimeter Institute for hospitality during the early stages of this project, and the Kalahari Meerkat Project for hospitality in the Kuruman River Reserve, South Africa, during its completion.



\end{document}